\documentclass{appolb}
\usepackage{graphicx}
\usepackage{array,amssymb,amsmath,booktabs}

\begin{document}
\eqsec  
\title{LARGE-$N$ PION SCATTERING, FINITE-TEMPERATURE EFFECTS AND THE RELATIONSHIP OF THE $f_{0}(500)$ WITH CHIRAL SYMMETRY RESTORATION%
\thanks{Presented at \textquotedblleft Excited QCD 2016\textquotedblright, Costa da Caparica, Portugal, March 6-12, 2016.}%
}

\author{Santiago Cort\'es$^{1}$, \'Angel G\'omez Nicola$^{2}$, and John Morales$^{3}$
\address{$^{1}$Departamento de F\'{\i}sica, Univ. de Los  Andes, 111711 Bogot\'a, Colombia \\
$^{2}$Departamento de F\'{\i}sica Te\'orica II, Univ. Complutense, 28040 Madrid, Spain \\
$^{3}$Departamento de F\'{\i}sica, Univ. Nacional de Colombia, 111321 Bogot\'a, Colombia.
}
}

\maketitle

\begin{abstract}
In this work, we review how the mass and the width of the $f_{0}(500)$ pole behave in a regime where temperature is below the critical chiral transition value. This is attained by considering a large-$N$ $O(N+1)/O(N)$ invariant Non-Linear Sigma Model (NLSM) such that we can study the dynamical generation of a $f_{0}(500)$ resonance. Introducing thermal effects via the imaginary time formalism allows us to study the behavior of the pole and relate it to chiral restoration.
\end{abstract}
  
\section{Introduction}

As lattice simulation results show \cite{Aoki:2009sc,Bazavov:2011nk}, analyzing low-energy phenomena as chiral symmetry restoration is needed for a proper description of the hadronic matter created in relativistic heavy ion collisions experiments (such as LHC-ALICE). Here we review a scenario \cite{CortesNicolaMorales} where a set of massless large-$N$ Nambu-Goldstone bosons \cite{Dobado:1994fd} interact with themselves and dynamically generate a scalar resonance (the $f_{0}(500)$), thus breaking the chiral symmetry, which should be restored when considering a thermal bath below the critical value; after attaining this, and since the pions do not gain thermal masses, we obtain that the chiral restoration is a second-order phase transition.

\section{Ellastic pion-pion scattering}

\subsection{Zero-temperature Regime}

We begin by considering the following $O(N+1)/O(N)$ Nonlinear Sigma Model \cite{Dobado:1994fd} with a metric and its associated vacuum constraint given by

\begin{align}
&\mathcal{L}_{NLSM}=\frac{1}{2}g_{ab}(\pi)\partial_{\mu}\pi^{a}\partial^{\mu}\pi^{b},
\label{NLSM} \\
&g_{ab}(\pi)=\delta_{ab}+\frac{1}{NF^{2}}\frac{\pi_{a}\pi_{b}}{1-\pi^2/NF^{2}}. 
\label{metric} \\
&f_{\pi}^{\,2}=NF^{2}.
\label{constraint}
\end{align}

\begin{figure}[htb]
\includegraphics[width=12cm]{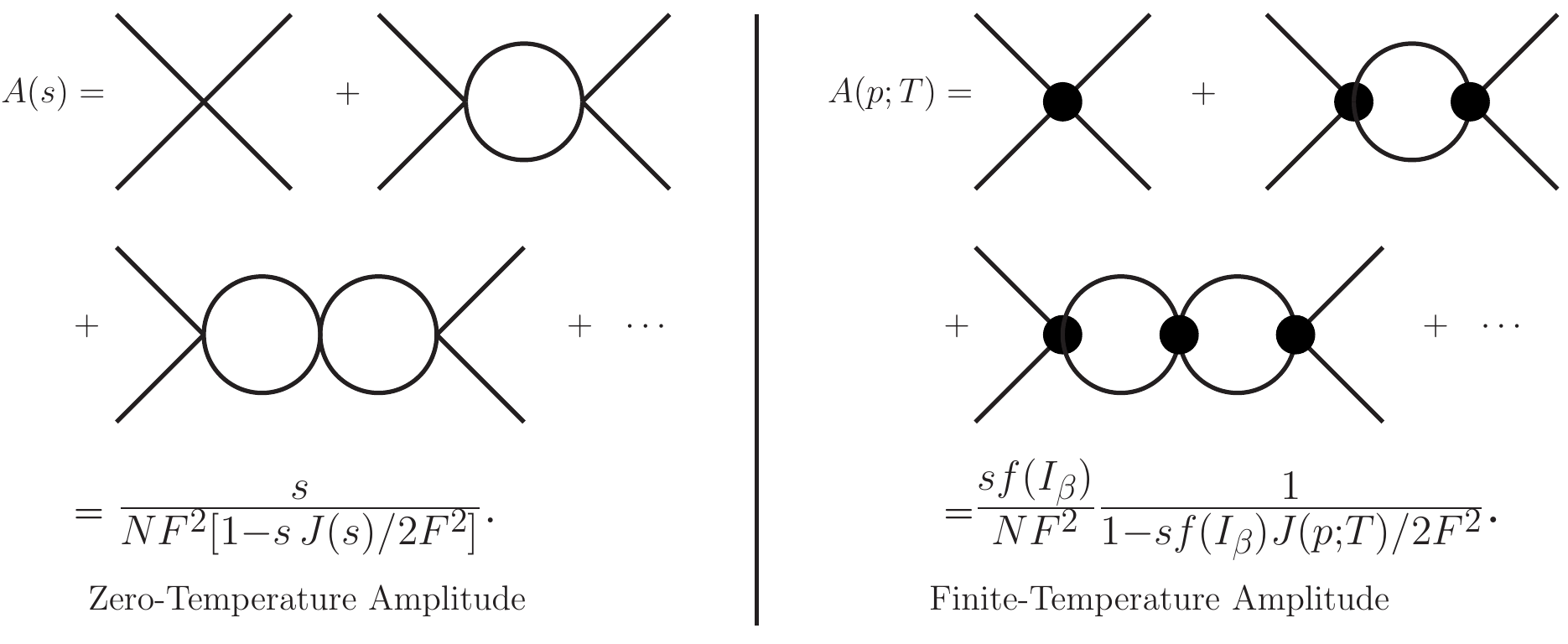}
\caption{Zero-Temperature and Finite-Temperature scattering amplitude for massless pions. The black circle represents the effective thermal vertex.}
\label{AmpZeroT}
\end{figure}

\noindent After expanding the non-diagonal term in (\ref{metric}) and reminding that we only want to study elastic scattering processes, we obtain the Feynman diagram and rule given in the left side of Fig \ref{AmpZeroT}, whose loop integral in the Dimensional Regularization scheme reads \cite{Gasser:1983yg}:

\begin{align}
&J(s)=J_{\epsilon}(\mu)+\frac{1}{16\pi^{2}}\ln\left(\frac{\mu^{2}}{-s}\right), \\
&J_{\epsilon}(\mu)=\frac{1}{16\pi^{2}}\left[\frac{2}{\epsilon}+\ln 4\pi-\gamma-\ln \mu^{2}\right]+\mathcal{O}(\epsilon).
\label{divergence}
\end{align}

\noindent Its proper renormalization is attained by redefining the 4-pion vertex as \cite{Dobado:1992jg}

\begin{equation}
\frac{s}{NF^{2}}\rightarrow\frac{s}{NF^{2}}G_{0}(s),\,G_{0}(s)=1+\sum_{k=1}^{\infty}{g_{0,\,k}\left(\frac{s}{F^{2}}\right)^{k}}.
\end{equation}

\noindent After considering this, we will absorb the divergence (\ref{divergence}) in the bare function $G_{0}(s)$ as $G_{R}^{\,-1}(s,\mu)=G_{0}^{\,-1}(s)-sJ_{\epsilon}(s)/F^{2}$, where $G_{R}(s,\mu)$ is written as a series expansion of a set of renormalized low energy constants $g_{R,\,k}(\mu)$. Then, the amplitude reads

\begin{equation}
A_{R}(s)=\frac{s}{NF^2}\frac{G_{R}(s;\mu)}{1-\frac{s\,G_{R}(s;\mu)}{32\pi^{2}F^{2}}\ln\left(\frac{\mu^{2}}{-s}\right)}.
\label{renormamp1}
\end{equation}

\noindent The partial wave associated to the scalar channel $I=J=0$ in the large-$N$ limit is given by

\begin{equation}
a_{00}(s)=\frac{1}{64\pi}\int_{-1}^{1}{NA_{R}(s)P_{0}(\cos\theta)d(\cos\theta)}.
\label{partialw}
\end{equation}

\begin{figure}[htb]
\centerline{
\includegraphics[width=12cm]{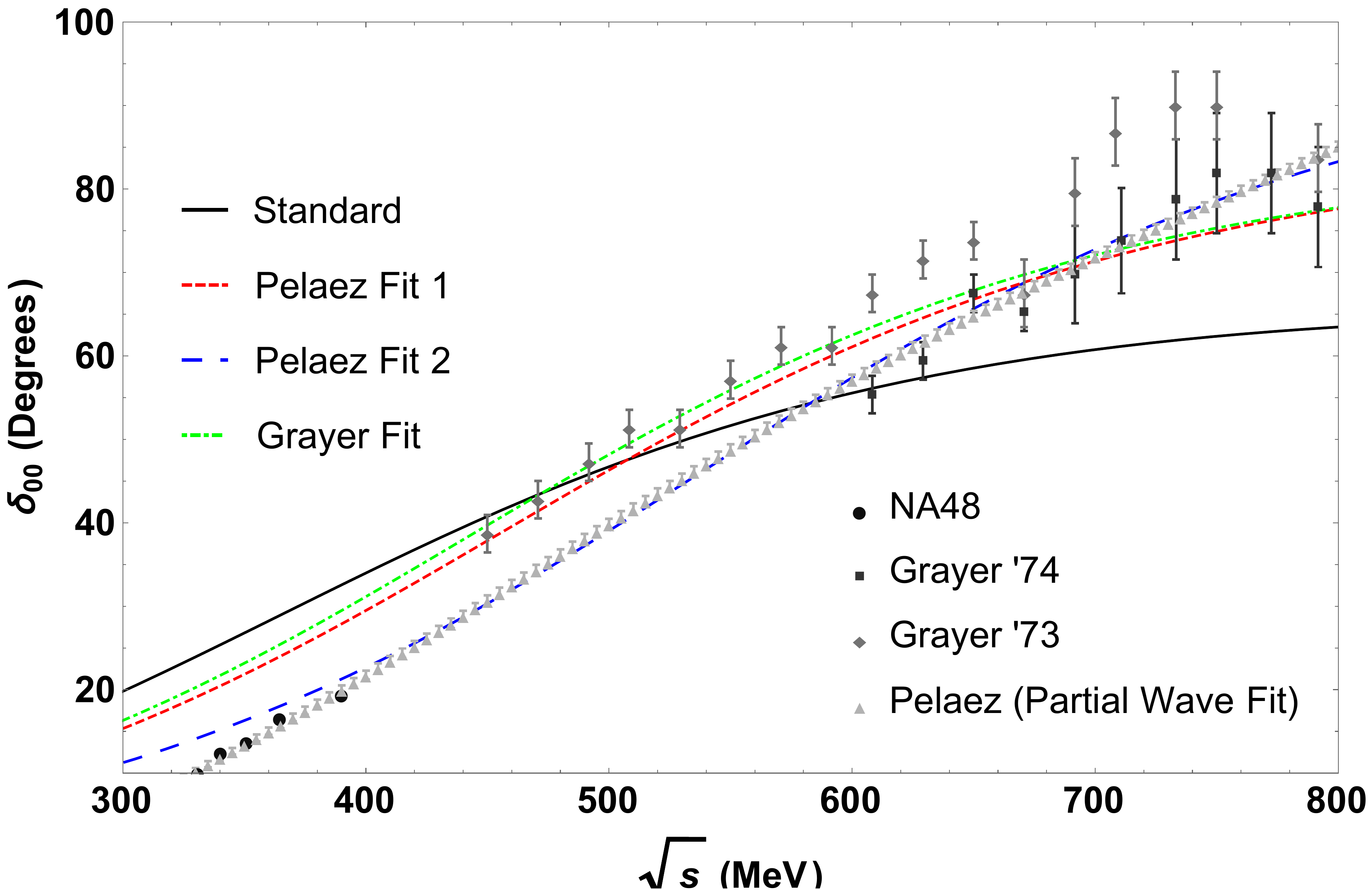}}
\caption{Partial wave fits to the $I=J=0$ scalar channel. References for the fit data are found in \cite{CortesNicolaMorales}, whereas the standard values are found in \cite{Dobado:1994fd}.}
\label{Delta00}
\end{figure}

\noindent This can be fitted to a proper set of data (both experimental and phenomenological) after choosing a scale compatible with $g_{R,\,k}(\mu)=0$. The results are shown in Fig. \ref{Delta00}, and the parameters are listed in Table \ref{table:fittable}. We do not take into account data neither close to the first threshold (where the pion mass matters) nor above 800 MeV (since strangeness is not considered).

\begin{table}
\begin{tabular}{c|ccc} \hline \hline
	Parameters & Grayer & Pel\'aez 1 & Pel\'aez 2 \\ \hline
	$F\pm\Delta F\text{ (MeV)}$  & 63.16$\pm$1.62 & 65 (fixed) & 75.98$\pm$ 0.16\\ 
	$\mu\pm\Delta\mu\text{ (MeV)}$ & 1523.35$\pm$143.34 & 1607.89$\pm$3.62& 2763.51 $\pm$ 23.81 \\ 
	$R^{2}$ & 0.9958 & 0.9951 & 0.9999  \\ \hline \hline
\end{tabular}
\caption{Parameters for the Grayer and Pel\'aez data fits  and their respective coefficients of determination.}
\label{table:fittable}
\end{table}

\subsection{Finite-temperature Regime}

The whole scattering process (taking into account effects of the thermal bath via the imaginary time formalism \cite{lebellac}) is given after building an effective thermal vertex that includes the contribution of all the powers of the tadpole $I_{\beta}=T^{2}/12$ that come from diagrams with 6 or more legs in the expanded metric (\ref{metric}). Taking this into account, we find an amplitude that reads as shown in the right side of Fig. \ref{AmpZeroT}, where $f(I_\beta)=(1-I_{\beta}/F^{2})^{-1}$ is the thermal tadpole function and the loop integral $J(p;T)=J_\epsilon (\mu)+ J_{fin}(p;T;\mu)$ has both zero and finite temperature contributions. We attain a proper renormalization of this quantity after redefining the vertices as

\begin{equation}
\frac{s}{(NF^{2})^{k+1}}\rightarrow\frac{s}{(NF^{2})^{k+1}}G_{0}^{\,k+1}(s),\,k=0,1,2,3,\cdots
\end{equation}

\noindent Furthermore, we find that the renormalized coupling $G_{R}(s,\mu)$ reads the same as in the zero-temperature case. Thus, the finite-temperature renormalized amplitude reads

\begin{equation}
A_R(p;T)=\frac{s G_{R}(s;\mu)}{NF^2}\frac{f[G_{R}(s;\mu) I_\beta]}{1-\frac{s G_{R}(s;\mu)f[G_{R}(s;\mu) I_\beta]}{2F^2}J_{fin}(p;T;\mu)}.\label{renampT}
\end{equation}

\section{The $f_{0}(500)$ Resonance and its Relationship with Chiral Symmetry Restoration}

\subsection{Thermal Unitarity}

After replacing the partial wave expansion (\ref{partialw}) into the the renormalized amplitude (\ref{renampT}), we can extract the imaginary part as

\begin{equation}
\text{Im}\left[\frac{1}{a_{00}(s+i0^+;T)}\right]=-\sigma_T(s,0),
\end{equation}

\noindent where $\sigma_T(s,0)=1+2n(\sqrt{s}/2)$ (here $n(x)$ is the Bose-Einstein distribution) is the thermal phase space for massless pions. This means that unlike previous perturbative results \cite{GomezNicola:2002an}, unitarity holds exactly in this framework.

\subsection{Mass and Width of the $f_{0}(500)$ Resonance}

\begin{figure}[htb]
\centering
\includegraphics[width=11.5cm]{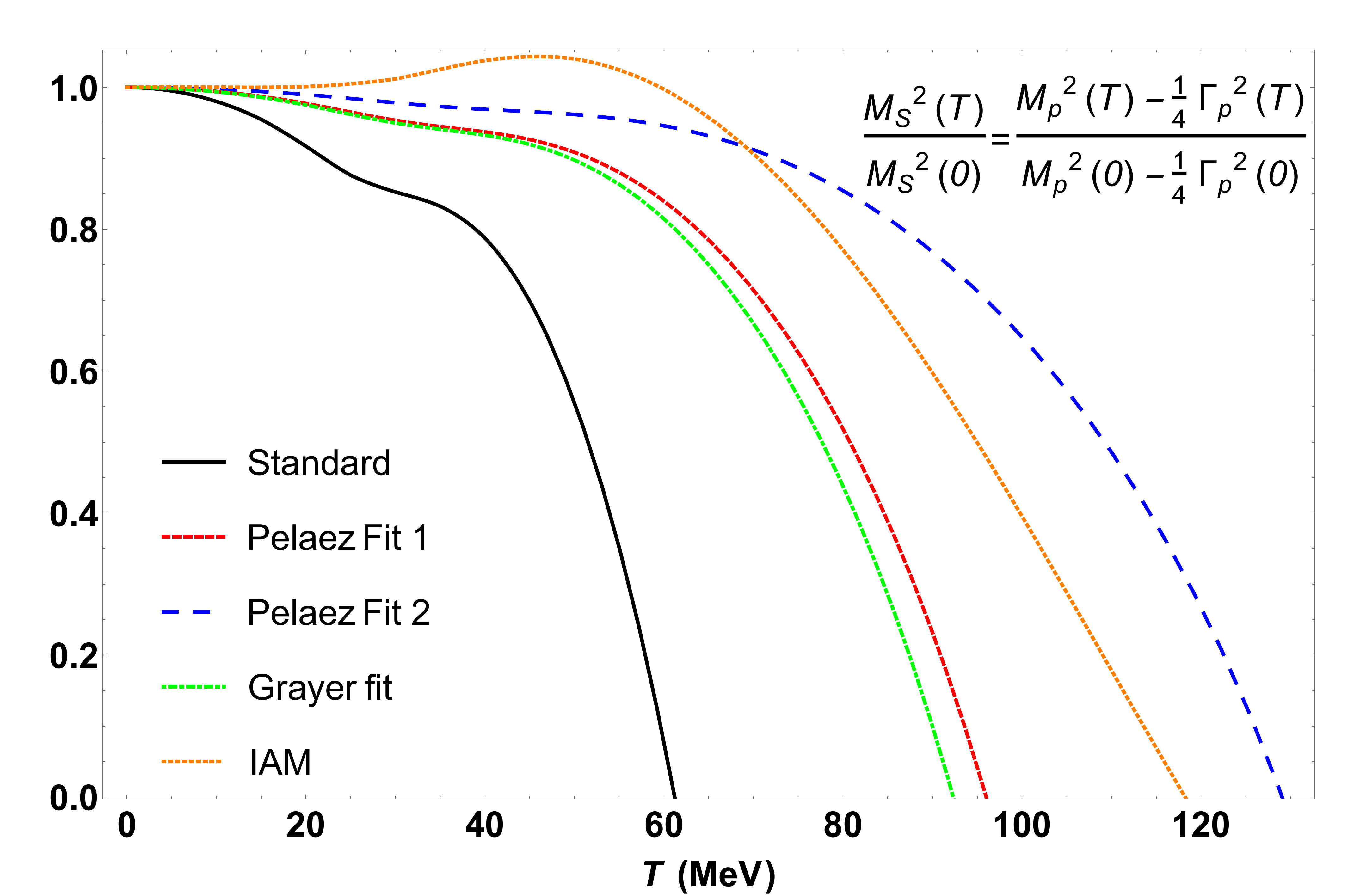}
\caption{Inverse scalar susceptibility as function of the temperature for different parameter sets, along with the IAM result. }
\label{scalarsuscep}
\end{figure}

Since unitarity was already checked, we can go to the second Riemann sheet and find the pole of $|a_{00}(s,T)|^{2}$; after doing this, we will have some insight about a symmetry-restoring behavior by studying the evolution with $T$ of the mass and the width of this resonance. We find a stronger evidence of this fact after obtaining the critical temperature and the critical exponent of the scalar susceptibility, whose $p=0$ limit is given as $\chi_{S}(T)\propto 1/\text{Re}\left\{s\right\}$ \cite{Nicola:2013vma}. Its unitarized form is such that $\chi_S(T)/\chi_S(0)=\left[M_p^2(0)-\Gamma_p^2(0)/4\right]/\left[M_p^2(T)-\Gamma_p^2(T)/4\right]$, whose inverse is plotted in Fig. \ref{scalarsuscep} along with the Inverse Amplitude Method result (IAM) \cite{Nicola:2013vma} for massless pions.

Table \ref{mainresults} lists our main results for the critical temperatures and pole positions, as well as the critical exponents and coefficients of determination for $\chi_{S}(T)$ \cite{CortesNicolaMorales}. We can see that our $T=0$ masses and widths can be compared with those given for a phenomenological fit and that they lie between the well-known experimental bounds (see \cite{CortesNicolaMorales} for details).

\begin{table}
\begin{tabular}{c|ccccc}  \hline \hline
\centering
	Fit & $T_{c}\text{ (MeV)}$ & $M_{P}(0)\text{ (MeV)}$ & $\Gamma_{P}(0)\text{ (MeV)}$ & $\gamma_{\chi}$ & $R^{2}_{\gamma_{\chi}}$\\ \hline
	Grayer & 92.33 & 438.81 & 536.47 & 0.875 & 0.99987 \\ \hline 
	Pel\'aez 1 & 96.00 & 452.42 & 546.26 & 0.938 & 0.99997\\  \hline
	Pel\'aez 2 & 129.07 & 535.53 & 534.59 & 0.919 & 0.99995 \\ \hline
	IAM & 118.23 & 406.20 & 522.70 & 1.012 & 1\\ \hline
	Standard & 61.20 & 356.97 & 566.05 & 0.842 & 0.99728 \\
	\hline \hline
\end{tabular}
\caption{Pole positions, critical temperatures and exponents and coefficients of determination of $\chi_{S}(T)$ for the fits considered so far.}
\label{mainresults}
\end{table}


\section{Conclusions}

Although we work in the Chiral Limit, our analysis of ellastic pion scattering at finite temperature in the large-$N$ expansion grants a description of the $f_0(500)$ pole thermal dependence that is consistent with previous works \cite{Nicola:2013vma}. Furthermore, the behavior of $\chi_{S}^{-1}(T)$ (saturated by the $f_0(500)$ pole) is consistent with a second-order phase transition, as seen in the lattice \cite{Bazavov:2011nk}. 

We have to point out that our $T_{c}$ results \cite{CortesNicolaMorales} are not far from the expected lattice values (about $\text{0.8}T_{\chi}$); besides, they are even closer to the result obtained for NJL-like models, i.e., $T_{c}\approx 100.7\text{ MeV}$ \cite{Berges:2000ew}. 

Our results for the critical exponents point out that they lie between the interval $\text{0.54}\leq \gamma_{\chi}\leq 1$, where the lower limit is given for an $O(4)$ three-dimensional Heisenberg model, and the upper limit is the exact result for a large-$N$ $O(N)$ four-dimensional nonlinear model (more details in \cite{CortesNicolaMorales}).

\medskip

\emph{We acknowledge financial support by Spanish research projects FPA2014-53375-C2-2-P and FIS2014-5706-REDT; Santiago Cort\'es acknowledges Univ. de los Andes and COLCIENCIAS.}

\end{document}